\documentclass[aps,12pt,a4paper]{revtex4-1}
\usepackage{textcomp}
\usepackage{graphicx}
\usepackage{amsmath}
\usepackage{wrapfig}
\usepackage[usenames]{color}
\usepackage{graphicx}

\begin{document}

\title{Influence of the substrate-induced strain and irradiation disorder on the Peierls transition in TTF-TCNQ microdomains}
\author{Vita Solovyeva, Anastasia Cmyrev, Roland Sachser and Michael Huth}
\affiliation{Physikalisches Institut, Goethe Universit\"{a}t,\\Max-von-Laue-Str. 1, 60438 Frankfurt am Main, Germany\\E-mail: levitan@physik.uni-frankfurt.de}

\begin{abstract}

The influence of the combined effects of substrate-induced strain, finite size and electron irradiation-induced defects have been studied on individual micron-sized domains of the organic charge transfer compound tetrathiafulvalene-tetracyanoquinodimethane (TTF-TCNQ) by temperature-dependent conductivity and current-voltage measurements. The individual domains have been isolated by focused ion beam etching and electrically contacted by focused ion and electron beam induced deposition of metallic contacts. The temperature-dependent conductivity follows a variable range hopping behavior which shows a crossover of the exponent as the Peierls transition is approached. The low temperature behavior is analyzed within the segmented rod model of Fogler, Teber and Shklowskii, as originally developed for a charge-ordered quasi one-dimensional electron crystal. The results are compared with data obtained on as-grown and electron irradiated epitaxial TTF-TCNQ thin films of the two-domain type.

\end{abstract}

\maketitle

\section{Introduction}
\label{sec:introduction}
As early as in 1955, Peierls demonstrated that a one-dimensional metal with a partially filled electronic band is unstable and undergoes a metal to insulator transition \cite{Peierls}. This transition is referred to as the Peierls transition. The Peierls transition is associated with the implicit perfect nesting of the Fermi surface in one-dimensional metals. It has gained additional significance with regard to the study of electronic correlation effects in low-dimensional systems and the interplay of correlation and disorder.

\begin{figure}[htb]
\includegraphics[width=0.8\textwidth]{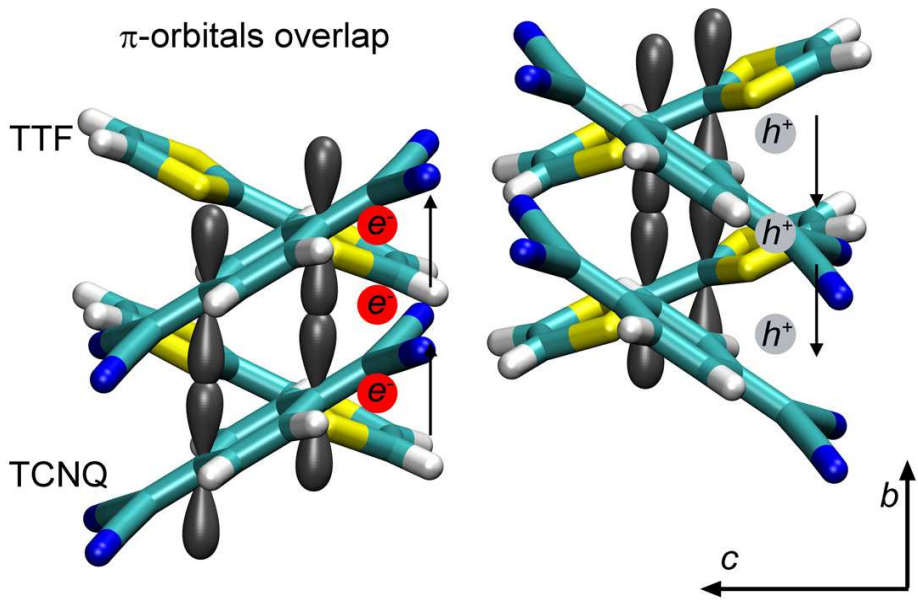}
\caption{Schematic illustration showing the overlap of $\pi$-orbitals for TTF and TCNQ molecules along the $b$ stacking direction. The VMD program was used for molecules representation \cite{VMD}.}
\label{fig:pi}
\end{figure}

The change of the Peierls transition temperature in  systems with reduced dimensionality, like  the organic charge transfer complex tetrathiafulvalene-tetracyanoquinodimethane (TTF-TCNQ),  with regard to biaxial strain and disorder  in the material was not previously studied due to some experimental difficulties. TTF-TCNQ is a quasi-one-dimensional organic metal which experiences a series of phase transitions at $T_H=54$~K, $T_I=49$~K and $T_L=38$~K \cite{Jerome_PRL_1978,Khanna,X_ray_scattering}, leading to a suppression of the metallic conductivity of the TTF and TCNQ chains and turning the material into a perfect insulator. The phase transition at 54~K is driven by a charge density wave (CDW) Peierls instability in the TCNQ chains \cite{Jerome_Schulz}. TTF-TCNQ consists of parallel segregated stacks of acceptor (TCNQ) and donor (TTF) molecules, where the stacks are aligned along the $b$-axis of the TTF-TCNQ as schematically shown in Fig.~\ref{fig:pi}. The overlap of $\pi$-orbitals arising along the stack direction causes a strongly anisotropic electrical conductivity.

The Peierls transition in TTF-TCNQ single crystals can be influenced by the presence of defects in the material and by hydrostatic pressure applied to the material \cite{Jerome_PRL_1978,Temperature-Pressure,Chu_pressure,Zuppiroli_irradiation,Chiang}.
The study of transport properties of TTF-TCNQ single crystals was extended to TTF-TCNQ thin films \cite{Chen_sapphire,TTF-TCNQ-halide,Indranil,Vita-TTF-TCNQ}. There are several aspects regarding the electronic properties of this material which can be studied in thin films that are not accessible in single crystals, such as substrate-induced strain and the effect of disorder caused by electron beam irradiation. Biaxial strain is very difficult to create in bulk single crystal, as is electron irradiation induced defects due to the small penetration depth of electrons.

In this paper we study the dynamics of low-lying charge excitations in TTF-TCNQ domains with large defect densities induced by electron irradiation, both, above and below the Peierls transition temperature $T_P$. Using electron irradiation for the induction of defects is advantageous for several reasons, one of which is the possibility to deposit locally large doses (reaching into the GGy regime) with very high lateral resolution and control within a scanning electron microscope.
Primary electron energies of several keV at a flux density of 250~As/m$^2$ have been employed. Due to the limited penetration depth of the electrons of about 400~nm at 5~keV the studies need to be performed on layers of TTF-TCNQ with thicknesses below about 200~nm in order to ensure a homogenous distribution of the induced defects. It is possible to grow epitaxial, $c$-axis oriented thin films of TTF-TCNQ on alkali-halide substrates \cite{Chen_sapphire,TTF-TCNQ-halide,Vita-TTF-TCNQ}. However, the epitaxy is of the two-domain type which inevitably leads to two complications. First, low-lying charge excitations originating from domain or grain boundaries will contribute to the measured transport properties. Considering the typical domain size for TTF-TCNQ, which is typically below 3~$\mu$m, this can only be avoided by high-resolution lithographic techniques. However, resist-based photo or electron-beam lithography is not readily applicable to organic charge transfer materials. Second, if the transport data are taken over multiple domains, a mixture ensues (series and parallel circuiting) of the contributions along the highly-conducting $b$-axis and the much less conducting $a$-axis. This renders data analysis much more complicated.

In this work we present and analyze temperature-dependent conductivity measurements and current-voltage characteristics taken on individual TTF-TCNQ domains with a typical size of 2.5~$\mu$m by 250~nm. These domains have been individually exposed to intensive  electron irradiation fluxes. We compare the results with data obtained on epitaxial, as-grown TTF-TCNQ films of the two-domain type. For the individual domains and irradiated thin film we find variable range hopping behavior of a different kind above and below about 50 K, corresponding to a slightly suppressed Peierls transition temperature, which is 54 K in bulk single crystals \cite{TTF-TCNQ-first}. We argue that the data below 50 K can be explained in the framework of the segmented metal-rod model of Fogler, Teber and Shklovskii developed for a quasi one-dimensional electron crystal \cite{Hopping-conduction2}. We speculate that the low-lying charge excitations above 50 K are mainly due to fluctuating charge density waves (CDW) which are subject to the same segmentation as assumed in the Fogler, Teber and Shklovskii model, albeit here an additional length scale enters, namely the average CDW-fluctuation correlation length. First experiments on the fabrication of TTF-TCNQ domains using a different approach than presented here were recently reported in \cite{nanowire_improved,Indian_nanowire_nanotech}.

\section{Experimental Methods}
\label{ssec:exper}

TTF-TCNQ domains (also called microcrystals in the following) were prepared by the physical vapor deposition method \cite{Mein_BEDT-TTF-TCNQ,arxiv} from as-supplied TTF-TCNQ powder (Fluka, purity $\geq$97.0$\%$) at a background pressure of $8\times10^{-8}$~mbar. The material was sublimated from a low-temperature effusion cell using a quartz liner at a cell temperature of 88~$^{\circ}$C.  The cell temperature was measured by a Ni-NiCr-thermocouple thermally coupled to the heated body of the effusion cell by copper wool. The substrate was kept at room temperature. The distance between the substrate and the effusion cell was 10~mm.

\begin{figure}[!htb]
\begin{center}
\includegraphics[width=0.9\textwidth]{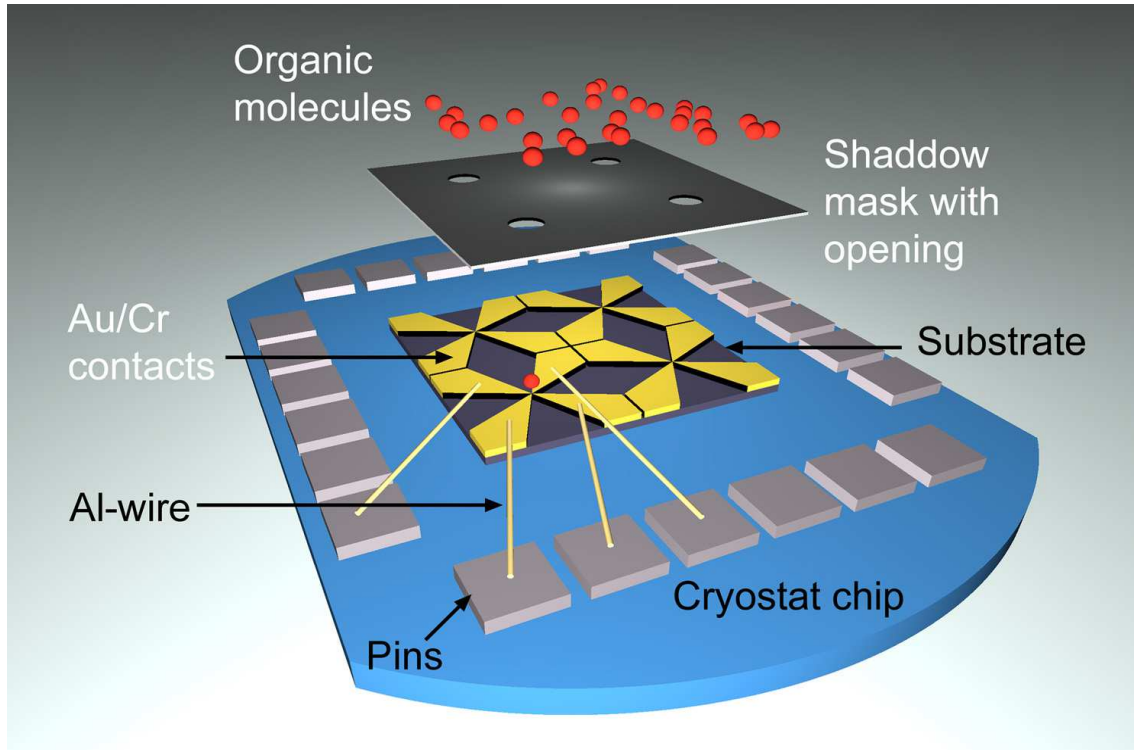}
\caption{Schematic illustration of the deposition process of the TTF-TCNQ domain on the substrate through the shadow mask. The substrate with bonded Au/Cr contacts is shown on the chip used for $in$ $situ$ conductivity measurements in the electron microscope during the irradiation experiment, as well as for the low-temperature measurements.}
\label{fig:deposition-process}
\end{center}
\end{figure}

As-supplied chemically cleaned Si(100)/SiO$_2$(285~nm) substrates were used in the experiments. The pre-patterned Au(50~nm)/Cr(20~nm) contacts were formed on the substrate by a
lithographical lift-off process and sputtering before the domain deposition. A shadow mask was used to define the region for TTF-TCNQ domain growth on the substrate between the metallic electrodes. One selected microcrystal on each substrate was contacted. The size of a typical microcrystal used in the measurements varied within the range of 1.5$\ldots$3~$\mu$m in length, 110$\ldots$550~nm in width and 10$\ldots$110~nm in thickness. The process of the domain formation is illustrated in Fig.~\ref{fig:deposition-process}.

Scanning electron microscopy (SEM) snapshots were done with a FEI xT Nova NanoLab 600 setup at a voltage of 5~  kV, beam current of 1.6~nA and dwell time of  3~$\mu$s. The thickness of TTF-TCNQ domains was determined from atomic force microscopy (AFM) measurements performed with a Nanosurf easyScan 2 AFM in non-contact mode.

For electrical transport measurements the microcontacts between the TTF-TCNQ domain and the pre-patterned Au/Cr contacts were fabricated by the combination of focused electron (FEBID) and ion beam induced (FIBID) deposition with trimethyl-methylcyclopentadienyl-platinum (MeCpPt(Me)$_3$) and tungsten hexacarbonyl (W(CO)$_6$) as precursor gases \cite{Huth-W,Huth-W2,Pt-Porrati}. Hereby the electrical contacts on top of the TTF-TCNQ microcrystal were fabricated by FEBID and then extended by FIBID. The width of the contacts formed in the area of the domain was 100~nm and the thickness was 40~nm. The electron beam was operated at 5~kV and 1.6~nA. The focused ion beam was operated near the microcrystal at 30~kV with a beam current of 10~pA. The microcontacts on the TTF-TCNQ domain were covered by a highly resistive Si(O) passivation layer using neopentasilane (Si$_5$H$_{12}$) as precursor by means of FEBID. The thickness of the layer was about 10~nm.

The electrical conductivity measurements were carried out in a $^4$He cryostat with a variable temperature insert allowing to cool down the sample from room temperature (300~K) to 4.2~K. A four-probe scheme for the transport measurement was applied.  The electrical transport properties were measured along the crystallographic $b$-axis of the TTF-TCNQ microcrystals.
Temperature-dependent resistivity measurements were performed at fixed bias voltage of 0.2~V corresponding to an electric field of 8000~V/cm. This rather large excitation level was needed to generate sufficient current. It is also related to the enhanced threshold voltage in the domain, as detailed in the next section. Cooling-heating cycles of the samples were repeated several times and the conductivity showed no indication of either hysteretic behavior or thermal-stress induced damage formation in the domains.

\section{Results}
\label{ssec:Res}

\begin{figure}[htb]
\includegraphics[width=1\textwidth]{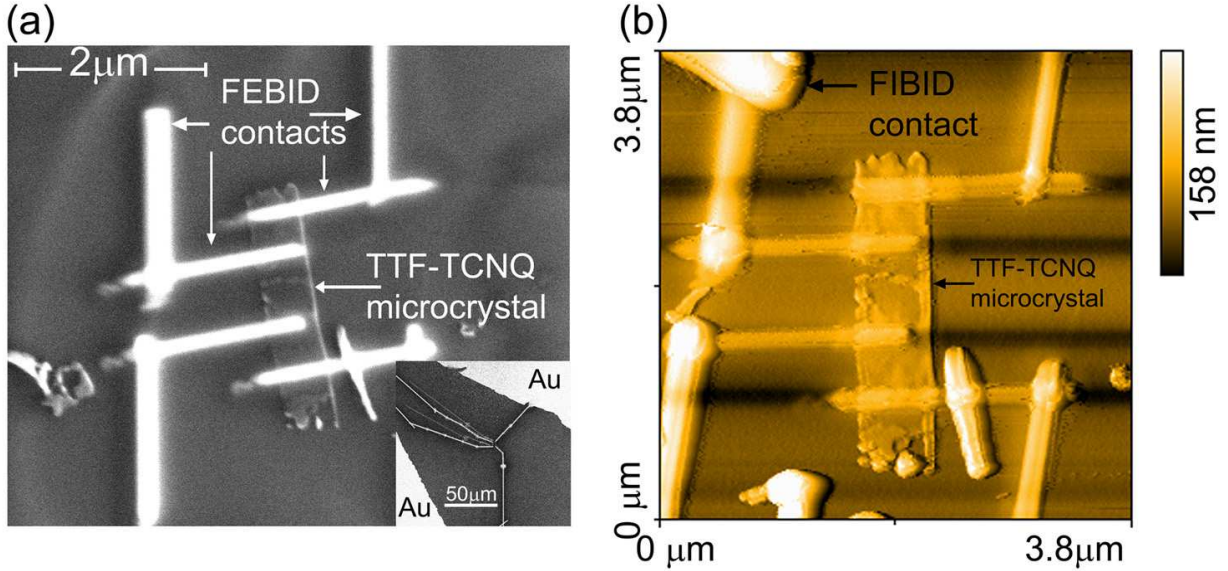}
\caption{(a) Exemplary SEM image of a TTF-TCNQ domain with contacts used for further analysis. The inset shows a zoomed-out SEM image of the TTF-TCNQ domain with FEBID/FIBID contacts together with the pre-patterned gold contacts. (b) AFM image of the TTF-TCNQ microcrystal with contacts. The thickness of the microcrystal in this example is about 15~nm. The AFM image was taken after the cryostat measurements.}
\label{fig:SEM}
\end{figure}

Figure~\ref{fig:SEM} shows a typical example of a TTF-TCNQ domain (microcrystal) with fabricated contacts. The length of the crystal in this
example is 2.5~$\mu$m and corresponds to the crystallographic $b$-axis, the width is 530~nm and the thickness is 15~nm (parallel to $c$-axis). Each domain was irradiated with the electron beam. The density of the electron flux was about 250~As/m$^2$. The dose for the electron radiation in the experiments ranged from 1.3~GGy to 2.4~GGy, as was determined by  Monte-Carlo simulations using the Casino computer program~\cite{Casino}. The relatively high irradiation dose stems from the small volume of the irradiated sample. High irradiation doses result in the formation of defects in the TTF-TCNQ domains. The energy which is needed to introduce one defect in TTF-TCNQ single crystals is 21.2~keV as reported in \cite{defect_measure} for X-ray radiation. We assume the same radiation efficiency for electrons in our case and calculate from the energy absorbed in a typical domain defect densities in the range of (0.62$\ldots$1.13)$\times10^{19}$~cm$^{-3}$.

The room temperature  electrical conductivity along the long axis of the microcrystal (the crystallographic $b$-axis of TTF-TCNQ) is in the range of 10$\ldots$30~($\Omega$cm)$^{-1}$. This value is more than one order of magnitude smaller than for high-quality TTF-TCNQ single crystals, which typically show $\sigma_b\sim$500~($\Omega$cm)$^{-1}$ \cite{conductivity_ratio}.
This reduced conductivity is caused by the electron beam irradiation of the TTF-TCNQ domain.

\begin{figure}[htb]
\begin{center}
\includegraphics[width=0.7\textwidth]{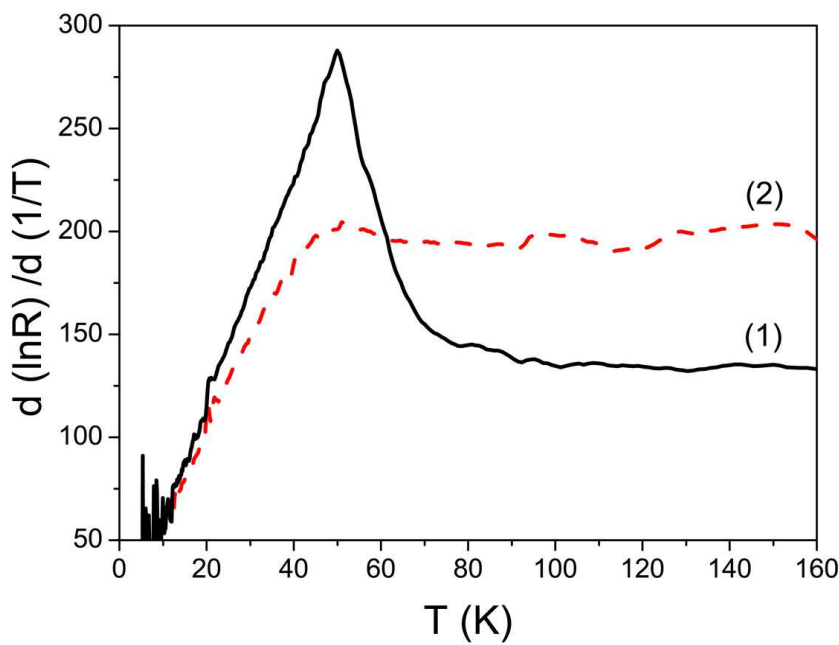}
\caption{Logarithmic derivative of the temperature-dependent resistance used to define the Peierls transition for (1) non-irradiated, and (2) irradiated TTF-TCNQ thin films. The irradiation dose was 1.4~GGy.}
\label{fig:irrad-film}
\end{center}
\end{figure}

The electrical resistivity of TTF-TCNQ domains fabricated with combined FEBID and FIBID processes shows a variable range hopping (VRH) behavior starting from room temperature down to low temperature \cite{Efros}:

\begin{figure}[htb]
\includegraphics[width=1\textwidth]{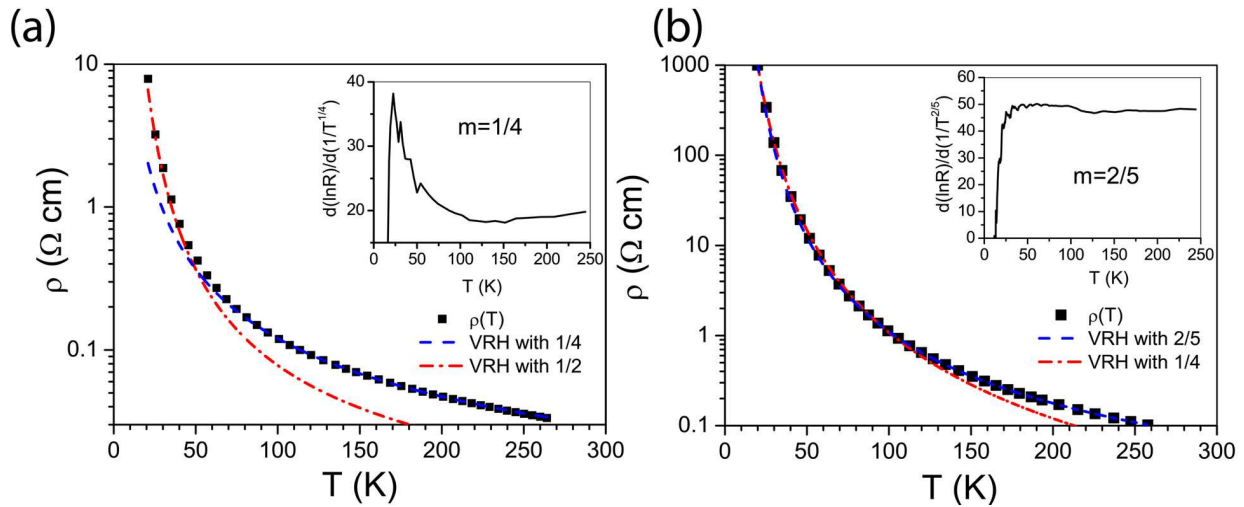}
\caption{Temperature dependence of the resistivity (squares)  measured for two TTF-TCNQ domains irradiated by electron beam with (a) 1.3 GGy, and (b) 2.4 GGy  radiation dose, respectively. The exponents $m$ used in Eq.~(\ref{eq:VRH}) are given for each graph. The applied voltage is 0.2~V. The insets show the temperature dependence of the logarithmic derivative $d(\ln R)/d(1/T)^m$ with $m$ as indicated.}
\label{fig:resistace-T}
\end{figure}

\begin{equation} \label{eq:VRH}
 R(T)=R_0\exp\left(\dfrac{T_0}{T}\right)^m,
\end{equation}

\noindent
where $T_0$ and $R_0$ are constants, $m$ is the exponent. In the experiment the exponent $m$ varies for different irradiation doses and experiences a change at a temperature of about 50~K as is shown in Fig.~\ref{fig:resistace-T}.
The values of the exponent for the temperature range above and below 50~K are summarized in Tab~\ref{tab:exponent}.
For reference purposes we analyzed the behavior of an epitaxial TTF-TCNQ thin film grown on NaCl(100) irradiated with the similar irradiation dose (about 1.4~GGy). The comparison of the temperature dependence of the logarithmic derivatives of the resistance for non-irradiated and irradiated TTF-TCNQ thin films is presented in Fig.~\ref{fig:irrad-film}. The logarithmic derivative of the resistivity is generally used to determine the Peierls transition temperature \cite{Gruner}. The values of the exponent for the thin films are also included in Tab.~\ref{tab:exponent}. It is important to note that the values for the exponents $m$ are to be taken with some caution. The respective fitting ranges, above and below about 50~K, are rather small. Nevertheless, the exponents given in Tab.~\ref{tab:exponent}  created significantly better fits than other physically plausible choices taken from the set $m\in(1, \frac{1}{2}, \frac{2}{5},\frac{1}{3}, \frac{1}{4})$. Also, the crossover temperature of about 50~K is quite well-defined as this is the temperature where the fits for the high- and low-temperature data tended to deviate from the experimental values, respectively.

\begin{table}[htb]
\caption{Exponent $m$ and $T_0$  (K) characterizing the electrical transport behavior of TTF-TCNQ domains and  thin films at temperature (1) $T\geq50$~K and (2) $T<50$~K.}
\label{tab:exponent}
\begin{center}
\begin{ruledtabular}
\begin{tabular}{lllll}
Sample	&  Domain 1.3~GGy& Domain 2.4~GGy & Thin film 0~GGy & Thin film 1.4~GGy\\
\hline
 $m$, $T_0$ ($T<50$~K)		      & 1/2, 1225            &	1/4, 3992363	    & 1/4, 8503056 	            &2/5, 13995 \\

 $m$, $T_0$ ($T\geq$50~K)          & 1/4,  130321        & 2/5, 15962   &  1,  137              & 1, 197           \\

 $l$(\AA{})           & 15             & 8  & 2.3$\times10^3$  & 13 \\
\end{tabular}
\end{ruledtabular}
\end{center}
\end{table}

The manifestation of the Peierls phase transition in the TTF-TCNQ domains is smeared out in the resistivity-temperature
dependence shown in Fig.~\ref{fig:resistace-T}. The absence of a minimum in the resistivity at about 54~K, i.e. the typical
indication of the Peierls transition in TTF-TCNQ single crystals, is a result of the electron irradiation of the samples in the SEM, which induces defects in the TTF-TCNQ microcrystal. The thermo-activated behavior of the resistivity generally found for TTF-TCNQ
thin films \cite{Vita-TTF-TCNQ} also does not fit the resistivity-temperature dependence of the TTF-TCNQ domains. The irradiated TTF-TCNQ thin films still show Arrhenius behavior (see Fig.~\ref{fig:irrad-film}). We are led to assume that the VRH behavior is caused by an interplay of the high concentration of defects in the
domains which induce localized states near the Fermi level and the specific geometry of the individual domains.

For all cases presented in Tab.~\ref{tab:exponent} the transport regime changes at a temperature below 50~K. In order to explain the change of the VRH exponent the theoretical model suggested by Fogler, Teber and Shklovskii (FTS) in \cite{Hopping-conduction2} for a quasi-one-dimensional electronic system was used. At temperatures below the phase transition the electronic charge distribution of an assumed Wigner crystal in the material is divided by impurities into metallic rods. Between the rods the electronic transport is performed via variable range hopping. The transport regime in the material is impurity dependent and also changes with temperature. FTS introduce an average length of the segment of the chain which is not disturbed by impurities $l=1/(Na_{\perp}^2)$, where $N$ is the defect concentration and $a_{\perp}^2$ is the area per chain. While in the theoretical model chains of the same type are considered, in the case of TTF-TCNQ  two types of chains should be taken into account. The Peierls transition studied here refers to the TCNQ chains resulting in $a_{\perp}\approx$10.4~\AA{} \cite{cryst-str-TTF-TCNQ}.
For each sample the average length of the segment was calculated and is given in Tab.~\ref{tab:exponent}. For the non-irradiated epitaxial thin film sample the concentration of the defects was taken from our previous analysis on TTF-TCNQ thin films grown on various substrate materials \cite{Vita-TTF-TCNQ}. The average length of the segment is responsible for the transport regime in the quasi-one-dimensional metal. The variable range hopping exponent $m$ in Tab.~\ref{tab:exponent} can be parameterized as follow \cite{Hopping-conduction2}:

\begin{equation}
m=\frac{\mu+1}{\mu+d+1},
\end{equation}

\noindent
where $d$ is the dimensionality and $\mu=0, 1, 2$ according to the physical situation were suggested in \cite{Hopping-conduction2}. For $d=3$, $\mu=0$ corresponds to $m=1/4$ and Mott behavior, $\mu=1$ to $m=2/5$, and $\mu=2$ to $m=1/2$ and Efros-Shklovskii variable range hopping behavior. When $m=1$ the transport regime is thermo-activated.

The results for the exponents given in Tab.~\ref{tab:exponent}
can be explained fairly well by the framework of the metallic rods model formulated in \cite{Hopping-conduction2}. For temperatures below 50~K the Mott behavior describes the transport behavior of non-irradiated TTF-TCNQ thin films, while for irradiated thin films the average segment is shorter, therefore, the transport regime is changed and the value of the exponent equals 2/5. The individual TTF-TCNQ domains studied here were irradiated with comparable irradiation dose and the inevitable uncertainty in the $l$ estimation does not allow for a clear statement concerning a crossover in the transport regime, which was possible in the case of irradiated and non-irradiated TTF-TCNQ thin films. The sample irradiated with 1.3~GGy shows the Efros-Shklovskii behavior and the sample irradiated with 2.4~GGy  displays the Mott behavior at temperatures below 50~K. Both these regimes are probable for strongly irradiated quasi-one-dimensional electronic systems as follows from \cite{Hopping-conduction2}. In any case the strong electron irradiation of the TTF-TCNQ domains does cause a complete suppression of the simple Arrhenius behavior commonly observed in thin films.
Also, the observed crossover of transport regimes stemming from the change of the VRH exponent $m$ at temperatures of about 50~K for the measured TTF-TCNQ domains is ascribed to the presence of the slightly suppressed Peierls transition in the
system which is assumed to persist despite of the increased defect
concentration.

\begin{figure}[htb]
\includegraphics[width=0.9\textwidth]{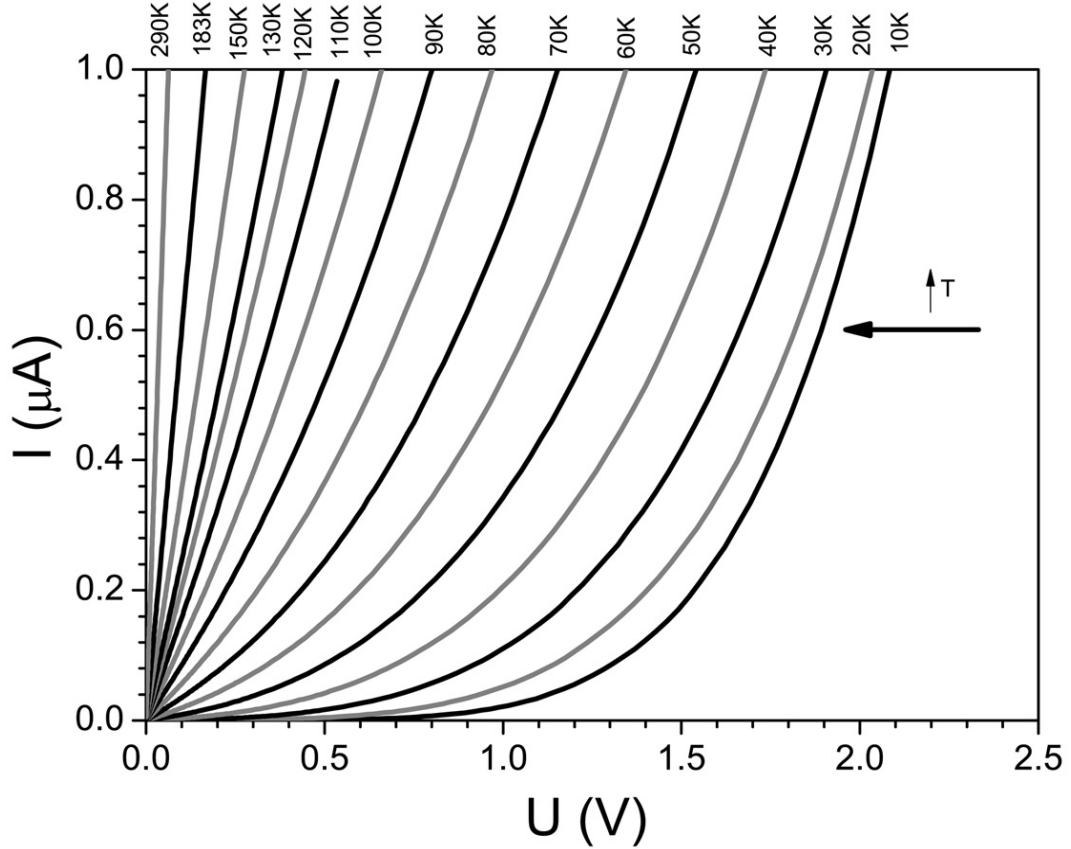}
\caption{Current-voltage characteristics for a TTF-TCNQ domain measured at various temperatures in the range of 10$\ldots$290~K. The irradiation dose for the microcrystal was 2.4~GGy.}
\label{fig:I-V}
\end{figure}

\begin{figure}[htb]
\includegraphics[width=0.9\textwidth]{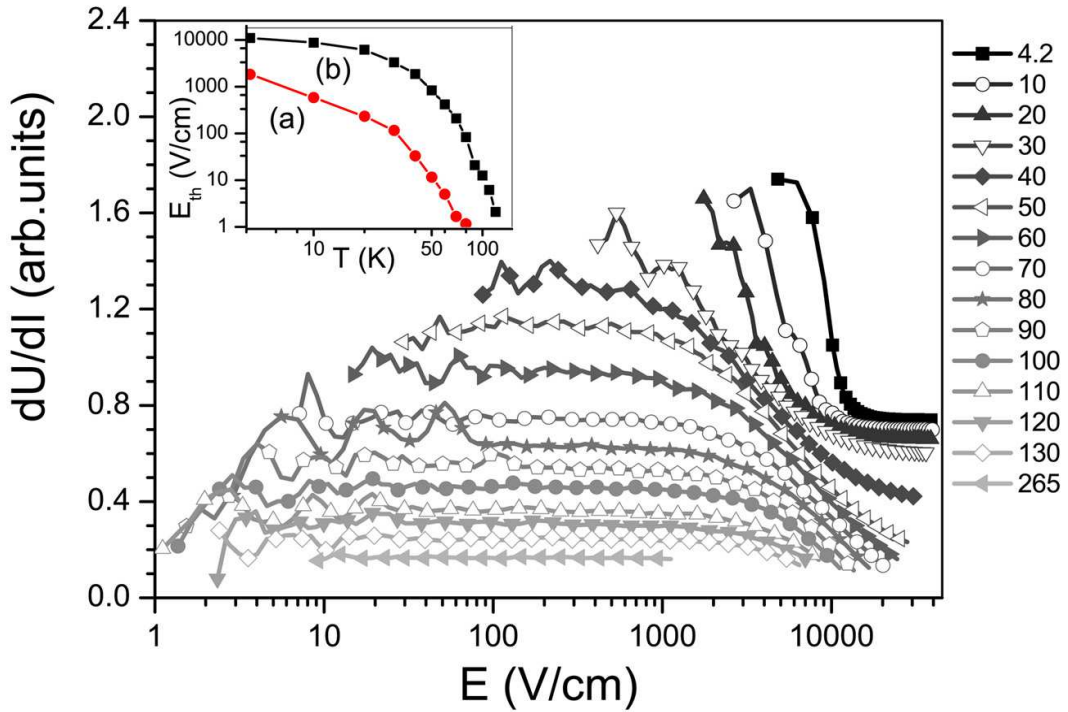}
\caption{Differential resistance of a TTF-TCNQ microcrystal measured in the temperature range of 4.2$\ldots$265~K. The irradiation dose of the microcrystal was 2.4~GGy. The inset shows the temperature dependence of the threshold electric field for the samples irradiated by an electron beam with (a) 1.3~GGy, and (b) 2.4~GGy, respectively.}
\label{fig:diff-res}
\end{figure}

In order to get more information about CDW contributions below $T_P$ (within the FTS scenario) and above $T_P$ (via CDW fluctuations) we performed current-voltage $I(U)$ measurements and analyzed possible threshold behavior. Several $I(U)$ curves for one domain measured at different temperatures are collected in Fig.~\ref{fig:I-V}. The $I(U)$ curves exhibit a non-linear behavior as the temperature is reduced to below about 100~K. The measured $I(U)$ characteristics were used to obtain the threshold  electric field for the TTF-TCNQ domain. In particular, Fig.~\ref{fig:diff-res}  shows the differential resistance derived from the data presented in Fig.~\ref{fig:I-V}. The threshold  electric field was defined from the dependence of the differential resistance on the electric field as the field where the differential resistance changes its behavior from constant, corresponding to Ohm's law, to a non-linear behavior. The inset to Fig.~\ref{fig:diff-res} shows the temperature dependence of the threshold electric field for the two investigated samples. The obtained threshold electric field at 10~K (800~V/cm and 8000~V/cm) is several orders of magnitude larger than commonly observed for TTF-TCNQ single crystals (10~V/cm \cite{E_threshold}) and for TTF-TCNQ thin films (3~V/cm). This increase of the threshold electric field is thought to be caused by two effects: (i) the relatively high concentration of defects in the domains \cite{E_threshold}, and (ii) the finite size of the sample, which has a severe influence on the threshold electric field in one-dimensional conductors as discussed in \cite{Zotov-UFN}.

At this stage we assume that the persistence of the threshold voltage at temperatures substantially higher than the Peierls transition temperature (50~K) is attributed to the pinning of fluctuating CDW regions which are known to dominate the charge transport in bulk crystals even far above $T_P$. If this CDW fluctuation components persist despite of the large defect density, it would arguably be subjected to the same segmentation of the chains as assumed in the
FTS-model \cite{Hopping-conduction2}, albeit here an additional length scale enters, namely the average
 CDW-fluctuation correlation length.

 As final point in our analysis we address an additional aspect which may be of relevance in developing a comprehensive understanding of the presented data. In \cite{Hiroshi_strain} the influence of a substrate-induced strain was studied on transport properties of the organic charge transfer
salt $\kappa$-(BEDT-TTF)$_2$Cu$[$N(CN$)_2]$Br. It was shown that by introducing either compressive or tensile strain
the superconducting ground state of the material can be dramatically changed from superconducting to insulating. Thus, substrate-induced effects need also  to be taken into account here. In our study we consider microscale TTF-TCNQ crystals deposited on top of a Si(100)/SiO$_2$ substrate.  The Si/SiO$_2$ substrate has negligibly small  thermal contraction \cite{Ibach} when compared to the thermal contraction of TTF-TCNQ \cite{thermoexpansion1}. Due to the difference of the thermal expansion properties of the substrate and the charge transfer complex a tensile biaxial strain is produced when the sample is cooled down. This does, of course, imply the assumption that the TTF-TCNQ microcrystal are fully clamped. A direct proof of clamping, as can in principle be provided by temperature-dependent X-ray diffraction experiments, is not feasible for individual microcrystals. However, due to the very small thickness of the microcrystals and in view of our results obtained of thin films \cite{Vita-TTF-TCNQ}, a clamped state is highly likely.

 A simple estimate \cite{X-ray} of this strain along the $a$ and $b$-axes of the TTF-TCNQ domain, with thermal expansion coefficients taken from \cite{thermoexpansion1} and elastic module from \cite{Young_module5_Fraxedas} at $T=50$~K, predicts the tensile strain and stress along the $a$-axis to be $\epsilon_a=0.7\%$ and $\sigma^{tens}_{a}=$0.4~GPa, and along the $b$-axis to be $\epsilon_b=2.3\%$ and $\sigma^{tens}_{b}=$1.3~GPa. TTF-TCNQ single crystals compressed by hydrostatic pressure of the same order of magnitude as calculated above experience a slight shift of the Peierls transition temperature of about $\pm4$~K \cite{Jerome_PRL_1978,Temperature-Pressure}. It is not proper to naively compare the possible effects of biaxial strain with results for hydrostatic pressure quantitatively. However, it may provide a qualitative idea about the range of changes of the Peierls transition temperature in the case of the biaxial strain as is expected to occur in the TTF-TCNQ microcrystals. The thermal tensile strain caused by the substrate-induced interaction decreases the overlap of the electronic wave function in TTF-TCNQ along the stack direction and one might rather expect an increase of the Peierls transition temperature. However, to our knowledge the influence of a biaxial strain on the Peierls transition temperature has not been studied by theoretical means so far. We are left with the qualitative statement that the strain effect is most likely not significant against the background of the change of the phase transition due to the induced defects.
The Peierls transition temperature in TTF-TCNQ domains irradiated with electrons is attributed to the point where the exponent in the variable range hopping behavior changes, i.e. at about 50~K. Due to the interplay of the impact from the defects (decrease of $T_P$) and from the biaxial strain (possible increase of $T_P$) we are not able to separate both effects in the present case.

\section{Conclusion}

In this paper the charge carrier dynamics in individual TTF-TCNQ domains fabricated by physical vapor deposition is discussed. The domain represents a system in which  thin film specific aspects like substrate-induced strain, size effects and disorder-induced changes in the electronic structure are combined. The results on TTF-TCNQ domains are compared with data obtained on epitaxial, as-grown TTF-TCNQ films of the two-domain type.

The contact fabrication process employed in the work is associated with defect incorporation which is caused by electron
irradiation of the organic microcrystals. The manifestation of the Peierls phase transition in the TTF-TCNQ domains is
smeared out in the resistivity-temperature dependence. The relatively high defect concentration results in a variable range hopping
behavior of the resistivity instead of commonly observed  metallic or thermo-activated one.
A non-linear behavior in the current-voltage characteristics develops below about 100~K. 

The properties of TTF-TCNQ domains differ dramatically from those of TTF-TCNQ single crystals and thin films. The influence of the substrate on the
grown domains is appreciable.
The development of tensile strain in the domain under cooling may lead to an
increase of the Peierls transition temperature. In contrast to this, the change of the Peierls transition temperature due to the developed biaxial strain is hindered by the impact from defects, which lead to a reduction of the phase transition temperature.
For the individual domains variable range hopping behavior of a different kind above and below a temperature of about 50~K, corresponding to a slightly suppressed Peierls transition temperature,
 was observed. The data below 50~K can be explained in the framework of the segmented metal-rod model of Fogler, Teber and Shklovskii  developed for a quasi-one-dimensional electron crystal \cite{Hopping-conduction2}. The observed threshold voltage for TTF-TCNQ domains is assumed to correspond to depinning of the charge density wave below $T_P$, which persists as a fluctuating contribution at temperature above $T_P$ despite the relatively large defect concentration.

The approach followed in this work provides a new pathway to the isolation of individual TTF-TCNQ (or other organic charge transfer) domains for studying size effects and clamping. The electron beam induced effects are very useful for studying the influence of irradiation induced defects. The combination of all of these aspects leads to additional complexities which cause us to conclude that (i) the combined action of finite size effects and disorder causes a crossover from Arrhenius-like to variable range hopping transport in TTF-TCNQ, (ii) the influence of biaxial strain has to be disentangled from the finite size and irradiation effects. It would be desirable to optimize the domain contact fabrication such that it can be performed without excessive defect formation. Work along these lines is in progress.

\label{ssec:concl}
\section{Acknowledgment}

The authors are grateful to the Sonderforschungsbereich/Transregio 49 project for financial support of this work. They also thank  Dr. Ilia Solov'yov for fruitful discussions and critical reading of the manuscript.

\end{document}